\documentclass[12pt]{article}

\newcommand{\forReview}{
  \renewcommand{\baselinestretch}{1.5}
  \addtolength{\parskip}{5mm}
  \addtolength{\oddsidemargin}{-5mm}
  \addtolength{\evensidemargin}{-5mm}
  \addtolength{\textwidth}{-5mm}
   \addtolength{\topmargin}{-18mm}
   \addtolength{\textheight}{30mm}
}
\forReview

\title{Molecular Motion on Semiconductor Surface via Tip-enhanced Multiple Excitation} 

\author
{
Tatsuya Momose$^1$, Ken-ichi Shudo$^{2,3}$, Hannes Raebiger$^{2}$, Shin-ya Ohno$^2$, \\ 
Takeshi Kitajima$^{4}$, Masanobu Uchiyama$^{3,5}$, Takanori Suzuki$^4$ \\
Masatoshi Tanaka$^2$\\
\\
\small{$^{1}$Graduate School of Engineering and $^{2}$Faculty of Engineering/Science }\\
\small{Yokohama National University, Tokiwadai 79-5, Hodogaya-ku Yokohama 240-8501, Japan}\\
\small{$^{3}$Riken, Hirosawa 2-1, Wako-shi Saitama 351-0198, Japan}\\
\small{$^{4}$School of Applied Sciences, National Defense Academy,}\\
\small{Hashirimizu 1-10-20, Yokosuka 239-8686, Japan}\\
\small{$^{5}$Faculty of Pharmaceutical Science, University of Tokyo,}\\
\small{Hongo 7-3-1, Bunkyo-ku Tokyo 113-0033, Japan}\\
}

\date{\today}

\usepackage[dvipdfmx]{graphicx}

\begin{document} 


\baselineskip24pt


\maketitle 

\newpage


\begin{abstract}
In a low-temperature study with a scanning tunneling microscope (STM), the irreducible lateral motion of a CO molecule adsorbed on a Si(001) surface showed a hyperlinear dependence on the tunneling current. This dependence implies that the adsorbate displacement is caused by multiple excitations of  adsorbate vibration modes, a situation thus far observed only at metal surfaces. The local vibronic temperature at the atomic scale on the surface heated by ohmic inelastic scattering of tunneling electrons indicates that there is an activation barrier of 0.11 eV for the irreversible motion of CO, in agreement with the adiabatic potential obtained from first-principles calculation. The highly efficient local heating is caused by a mid-gap state at the surface induced by the electric field of the STM tip.
\end{abstract}

\section{Introduction}

Techniques of atom-by-atom manipulation, including desorption, hopping, and chemical reaction, to modify surfaces at nanometer scale with scanning tunneling microscopes  (STM) are well-established \cite{Eigler1990,Chiravalloti2009,Shen1995}. High electron density of the tunneling current plays an important role in these surface reactions, which are typically interpreted in terms of two mechanisms. (i) In the Menzel-Gomer-Redhead (MGR) model, an electronic excitation to a stable anti-bonding state leads to nuclear motion along an adiabatic potential \cite{Menzel1964,Redhead1964}. This usually occurs  when the applied bias voltage is relatively high \cite{Persson1997}, and the reaction rate $R$ shows a linear dependence on the tunneling current $I$ \cite{Bartels1998,Stipe1997,Olsen2009}. (ii) In the ladder-climbing model \cite{Komeda2005}, energy quanta of a local vibration are multiply excited by the inelastic scattering of tunneling electrons applied at a relatively low bias voltage, until  the accumulated energy exceeds  the activation energy. In this case, the reaction is governed by the power-law $R \sim I^n$, where $n$ is the reaction order. 
Such multiply excited process have been observed chiefly on metal surfaces thus far 
\cite{Komeda2002,Stipe1998}, but for very limited cases \cite{Shen1995,Kirimura2006}.
Accumulation of vibrational energy at molecules on surfaces is competitive against 
dissipation into equilibrium.
For instance, the lifetime of the C-O stretching vibration on Pt(111) is 2.2 ps \cite{Germer1993},
whereas that of CO on Si(001) is 1.87 ns \cite{Lab2005,Lab2006}. 
The latter longer lifetime suggests that the 
accumulated energy would overcome the reaction barrer, but
the observed rate in vibration-induced reaction follows a linear relation on a semiconductor \cite{Zhu1992}.
Because the mechanism is still unclear, 
we focus in this work on the role of covalent electronic states localized near the  molecule adsorbed on the surface. 
Our detailed STM experiments combined with first-principles calculations reveal that the accumulated energy of CO vibrations on Si(001) during STM observation is sufficient to induce  induce a change of the adsorption structure of the CO molecules. We will show that the power law breaks down in this system, and that the vibrational temperature is a good indicator of atomic-scale reactions proceeding on the semiconductor surface.

	The Si(001)-$c$(4$\times$2) surface adsorbs CO molecules below $\sim$200 K without decomposition \cite{Bu1993,Young1995}. There are two stable adsorption sites \cite{Bacalzo1999,Imamura1998,Lu2002}, where CO either (i) terminates a dangling bond at the down-Si of the dimer (T-CO), or (ii) sits on the metastable bridge site of two Si atoms of a dimer (B-CO).  On a clean surface, the formation of B-CO requires an activation energy of 1.1 eV, which can be achieved, for example, by an accelerated CO molecular beam \cite{Hu1997}. In our STM images of the CO-adsorbed surface, we show that bright spots emerge after scanning, which indicates that B-CO is formed in response to the tunneling current. We discuss this process in terms of an inelastic tunneling mechanism,  based on density functional calculations.

\section{Experimental and Computational Details}
We used on an ultrahigh vacuum (UHV) chamber with a base pressure of less than $5 \times 10^{-9}$ Pa, equipped with an STM (Omicron VT-STM). Our STM tip was made by electrochemical etching of a polycrystalline tungsten wire and further cleaned by electron bombardment. 
Before CO exposure, a clean $c$(4$\times$2) structure of  Si(001) ($n$-type, 0.02 $\Omega\cdot$cm)  was confirmed by STM observation, and the sample was cooled to $\sim$90 K. The clean surface was exposed to CO at a level of more than 100 Langmuir \cite{Langmuir}, and STM observation was performed at various tunneling currents. To model the CO adsorption, density functional calculations were performed with the VASP code \cite{Kresse1996,Kresse1999}.
The energy cutoff of the plane wave basis set was set at 400 eV. We modeled the surface of a (2$\times$3) unit cell as containing eight layers of Si separated by more than 10 {\AA} of vacuum. CO molecules are adsorbed at the reconstructed top layer, while the dangling bonds in the unreconstructed bottom layer are terminated with hydrogen atoms. For the Brillouin-zone integration, a 4$\times$4  \textbf{k}-point grid was used \cite{Ramstad1995}. All atoms in the top seven layers were relaxed until the force on each atom fell below 0.02 eV/{\AA}. 

\section{Results}
Fig. 1 shows successive STM images of a 50$\times$50 nm$^2$ area after the CO exposure was stopped.
A few bright spots and darker islands were observed, as reported previously \cite{Yamashita2003}. 
We found that the number of bright spots increased upon repeated scanning even after the CO exposure was stopped. 
The seventh scan covered a wider area of 100$\times$100 nm$^2$, as shown in Fig. 1(e). 
Here, the number of bright spots is obviously increased in the marked area, which had already been scanned six times,
indicating that formation of the bright spots was associated with the repeated STM scanning. 
The bright spots did not appear on dimers where the adjacent dimer showed a bright spot.
Disappearance of the bright spots was rarely observed.

The magnified STM image in Fig. 2(a) shows that the bright spots are formed at the centers of the Si dimers, and thus can be interpreted as B-CO.
On the other hand, the T-CO structure was not clearly observed, suggesting that T-CO is invisible to STM at the applied bias, because the shape of the electron distribution is similar to that of a dangling bond of intact Si dimer.
A schematic structure of one B-CO and two T-CO in a 2$\times$3 unit cell is superimposed on images of the electronic isosurface mapping calculated using the Tersoff-Hamann model in Fig. 2(b) and (c). Fig. 2(b) was calculated for a charge neutral system ($N_e$),  and Fig. 2(c) for a system with one excess electron injected ($N_e+1$) due to the electric field effect induced by the bias between the sample and the STM tip \cite{Vanpoucke2010}. 
B-CO appears as a hemisphere in the experimental image. The calculated image of  $N_e$ clearly splits into the two lobes of a $p$-orbital, while that of $N_e+1$ is much closer to a hemisphere.
Among calculated systems with B-CO, only the configuration in which one B-CO is sandwiched by two T-CO pointing in opposite directions was stable after structural optimization. In all other cases, B-CO moved to the terminal site during the optimization due to the lack of an activation barrier against transformation from B-CO to T-CO. This is presumably why no adjacent B-COs were found in STM images after repeated scans.
The energy of  T-CO is lower than that of B-CO, which is consistent with the fact that  the number of B-CO is negligible in the experimental image of the first scan after CO exposure.
Thus, the emerging bright spots indicate a lateral displacement of CO, i.e., T-CO is initially formed during the gas exposure, and then transformed to B-CO due to the STM scanning.

The reaction rate ($R$) of displacements is plotted as a function of the tunneling current in Fig. 3. 
Because the rate is not proportional to $I$, the reaction cannot be explained in terms of the MGR model caused by a single electron transition. 
This suggests that the reaction is a multiple excitation process. However, 
as shown in the inset, we could not fit our experimental data with a simple power law ($R \propto I^n$). Therefore, we introduced a local vibrational temperature $T_v$ to describe the degree of vibration excitation so that the reaction rate can be expressed by an Arrhenius-like relation \cite{Gao1997}
	\begin{equation}
		R = A \exp({-\frac{E_a}{k_B T_v}})\;.
		\label{eq:fit-2}
	\end{equation}
Here $\hbar$ is Planck's constant, $E_a$ is the activation energy for the reaction, $k_B$ is Boltzmann's constant, and the pre-exponential factor $A$ is related to the vibrational excitation rate and the number of multiple excitations of vibrational quanta spaced by $\hbar \omega$. 
This equation leads to a simple power law in the case where the molecular orbitals have a large overlap with substrate states near the Fermi level on metallic surfaces \cite{Gao1997,Stroscio2004}.
In contrast to this frequently employed approximation, 
energy dissipation of tunneling electrons is described as $P = \gamma I^2$ \cite{Hliwa2002}, 
where $\gamma$ is the ohmic coefficient of local vibrational heating due to resistance at the tunnel junction. 
The vibrational temperature is given by $T_v = T_0 + \gamma I^2$, using the substrate temperature  $T_0$. 
Hence we obtain
	\begin{equation}
		R=Ae^{-\frac{E_a}{k_B (T_0 + \gamma I^2)}},
		\label{eq:fit-3}
	\end{equation}
which excellently fits the curve in the main panel of Fig. 3. 
Note that $A$ and $\gamma$ contain the thermal dissipation of energy and the frequency of inelastic scattering, respectively.
The obtained parameters are $A = 10^8{\sim}10^9$ s$^{-1}$, $\gamma = 14.3 \pm 7.3$ K/A$^2$, and $E_a = 0.11 \pm 0.05$ eV. 
The activation energy $E_a$ is associated with the molecular motion of CO. 
The calculated adiabatic potentials for the motion of CO along the [110] direction (Fig. 4(a)) 
for no additional charge ($N_e$) and one additional electron ($N_e+1$) are shown in Fig. 4(b) and (c), respectively. 
The calculated activation energy for the transformation from T-CO to B-CO is reduced from 0.17 to 0.15 eV when the tip-induced charging of the surface is considered, in good agreement with the experimental results. 

\section{Discussion}
Due to multiple excitations of local vibrations, T-CO transforms to B-CO with a potential barrier of 0.154 eV,
much smaller than the value previously calculated in a small cluster \cite{Hu1997},
indicating that repulsion between CO molecules on the surface (included in our model) is non-negligible. 
In fact, the barrier is small enough that three or four excitations  of vibrational quanta at T-CO overcome it.
The reaction rate $R$ obtained with STM does not fit with $I^n$ for any $n$, 
but is well explained by vibrational heating proportional to the square of tunneling current $I$.
This result also indicates that an increase of temperature by several tens of K is sufficient to induce this structural change.
Moreover, high-resolution electron energy loss spectroscopy shows no peak related to B-CO at $\sim$80 K \cite{Bu1993,Young1995}, while at higher temperature a shoulder of B-CO \cite{Hu1997} appears on a T-CO peak \cite{Young1995,Kubo1997}. 
This experiment suggests that the formation of B-CO can be induced at thermal equilibrium, 
but the STM heating is ultimately localized at the adsorbed site \cite{Komeda2002_2,Stipe1998}.
As there are many vibronic modes at the sites, the vibrational temperature is a good indicator for reactions.
Indeed, this idea is widely accepted for reactions by electronic excitation on metal surfaces \cite{Stroscio2004}.
In this inequilibrium model, the temperatures of electrons, the local vibration of adsorbate and the  substrate phonons after electronic excitation by photons reach equilibrium separately \cite{Bartels2004}. 
Our results show that introducing the vibrational temperature is also valid for surface reactions induced by STM current  \cite{Gao1995}. 
In our semiconductor case, the temperatures of CO and Si substrate are different under local current heating with the STM,
and the efficiency of this heating depends on the local structure at the atomic scale.
On a semiconductor surface, the vibrational energy dissipates into the bulk, and the dissipation is
dependent on the vibrational modes \cite{Sakong2008}.
The lifetime of the T-CO vibration is as long as 2.3 ns \cite{Lab2005}, which is comparable to the average
period of electron/hole injection in our experiment (an electron per ns, approximately).

The transformation from T-CO to B-CO is irreversible, which suggests that the temperature is not homogeneous 
in an area wider than the STM resolution, as homogeneous heating would lead to
a constant ratio of T-CO and B-CO.
An STM tip at a bias voltage of 1.6 eV introduces 1.3 electrons in a unit cell on Si
(electronic susceptibility of $\epsilon \simeq 12 \epsilon_0$) \cite{Hu1997},
corresponding to our calculation of the $N_e +1$ state. We find a surface localized state  in the band gap
for the T-CO structure. On the other hand, the extra electron in B-CO delocalizes to the bulk.
A surface localized state is known to make a large contribution to inelastic tunneling
in the case of the $\pi$-orbital in benzene adsorbed on Si(001) \cite{Alavi2000}, and thus we expect that heating at the T-CO site would occur more efficiently than at the B-CO site. 

The power law of surface chemical reactions has been successfully applied for metal surfaces, 
where excitation only within the adsorbed molecule is considered.  
Dissipation of energy at the metallic surfaces is very rapid, while the lifetime of molecular excitation is much longer.
In the STM-induced process on semiconductor surfaces, however,
the excitation at the adsorbate and dissipation into the substrate are competitive.  
The vibrational energy of B-CO, obtained in our calculation, is at resonance with the Si-bulk phonon continuum,
whereas that of T-CO is off the resonance \cite{Kress1968}. 
Therefore, the vibrational energy of B-CO is strongly coupled to the bulk phonons, and decays very rapidly.
The probability for transformation from B-CO to T-CO is expected to be extremely low due to 
the short lifetime of the B-CO vibration.

The mechanism of electron transfer is described in terms of hole injection, 
as shown schematically in Fig. 5.
The inelastic tunneling process at T-CO is very efficient because of 
the mid-gap state locally induced by the electric field $\mathbf{E}$ from the STM tip.
The potential near CO is changed by 
both the hole injection into the local state and its recombination with a donor state that follows.
During these transfer processes, energy of the order of $\sim$150 meV can be lost, 
which corresponds to the vibronic quantum. 
This inelastic scattering resonantly excites the local vibration involved in the CO movement.
At B-CO,  in contrast, the hole transfers into the substrate, where it immediately diffuses 
deep into the bulk without an inelastic process at the surface. 

\section{Summary}

T-CO adsorbed on a Si(001) surface transforms to B-CO during STM experiments due to electronic excitations.
The rate of this reaction does not follow a power law, but reflects multiple excitations of local vibration modes.
The local temperature, an indicator of locally accumulated energy at the atomic scale,
follows the ohmic relation $\propto I^2$.
In this process, the key feature is  vibronic excitation by inelastic scattering of an electron/hole tunneling
through the local state induced by the tip field at T-CO. The long lifetime of the CO vibration
allows multiple vibronic excitations to occur.
In a semiconductor system, we have thus demonstrated that 
adsorbate structure modification can be induced by atomic-scale heating effectively owing to 
the STM tip.

\section{Acknowledgment} 
Calculations were partly performed 
on the Riken Integrated Cluster of Clusters.
H.R. was funded by a Grant-in-Aid for Young Scientists (A) grant (No. 21686003) 
from the Japan Society for the Promotion of Science.


\clearpage


\begin{figure}[htbp] 
	   \centering
	   \includegraphics[width=14cm]{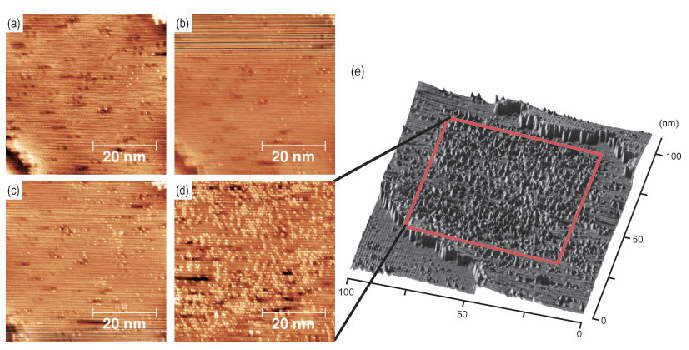} 
	   \caption{
(a-c) STM images obtained in the first to third scans of a CO-adsorbed Si(001) surface after CO exposure.
125 s was required to obtain each image. The scanned area is 50$\times$50 nm$^2$. (d) An STM image obtained at the seventh scan after cessation of CO exposure. (e) An STM image of a wider area (100$\times$100 nm$^2$) taken after the seventh scan to show the surrounding region.  All images were obtained with a sample bias voltage of -1.6 V and tunneling current of 200 pA. }
\end{figure}
		
\begin{figure}[htbp] 
	   \centering
	   \includegraphics[width=12cm]{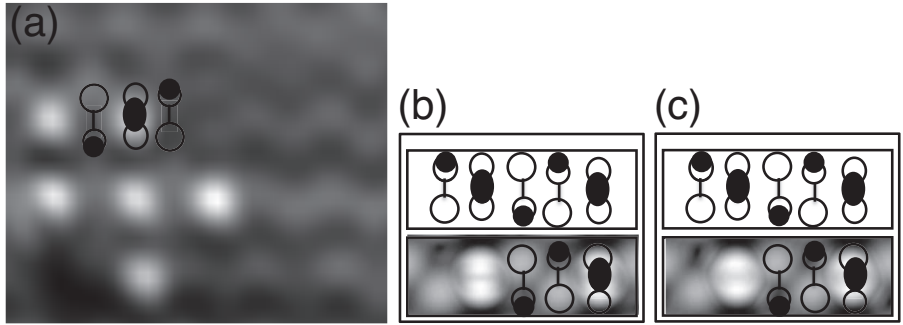} 
	    \caption{
 (a) An STM image showing CO adsorbed on Si(001) at -1.6 V.
 (b) and (c) show the simulated structures of T-CO and B-CO with $N_e$ and $N_e+1$ electrons, respectively.  
Filled and open circles represent CO molecules and Si atoms, respectively.
Below: the simulated images with the corresponding schematic superimposed. 
}
 \end{figure}

\begin{figure}[htbp] 
	   \centering
	   \includegraphics[width=10cm]{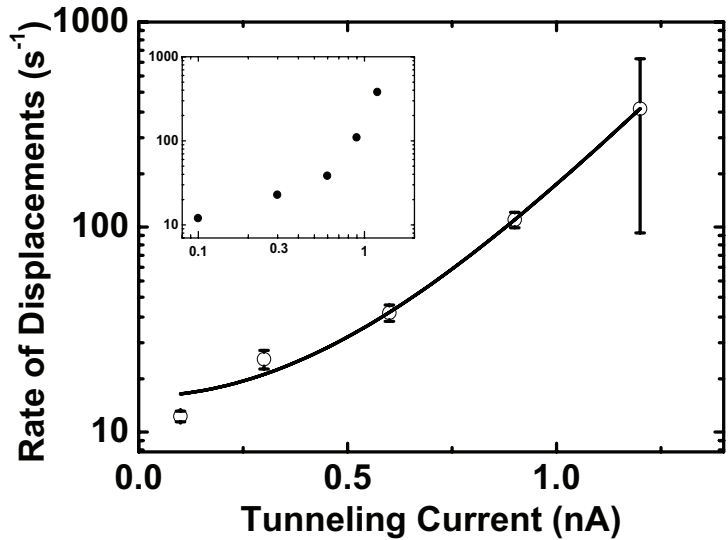} 
	   \caption{Frequency of emerging bright spots in STM images during the average residence time of the tip at a dimer. The tunneling current was set to 100, 300, 600, 900 pA or 1.2 nA. Inset is the same data but plotted in log-log scale to show that the points do not lie on a straight line (See text).
	   }
\end{figure}

\begin{figure}[htbp] 
	 \centering
	 \includegraphics[width=6cm,clip]{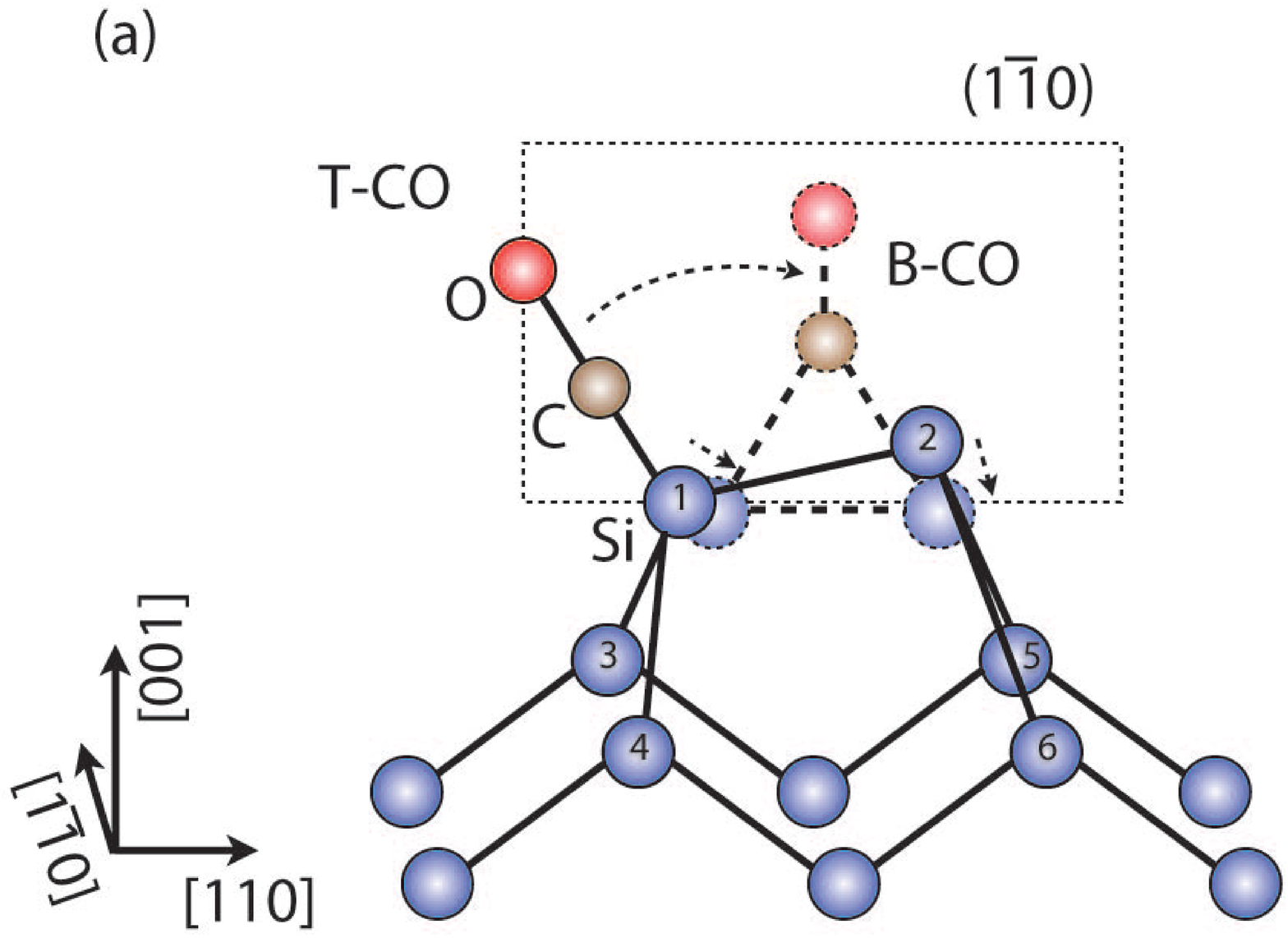} \\
	 \includegraphics[width=6cm,clip]{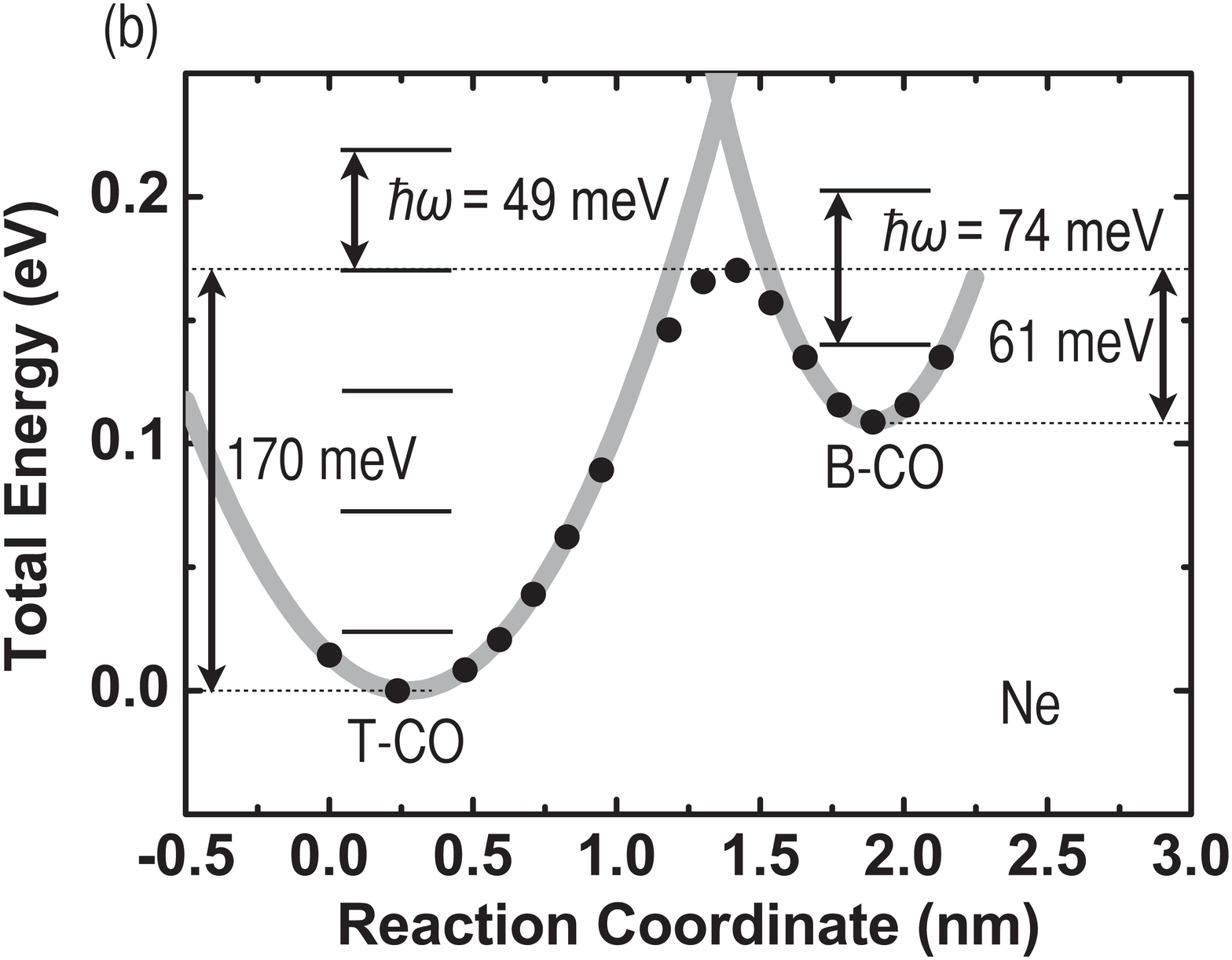} 
	 \includegraphics[width=6cm,clip]{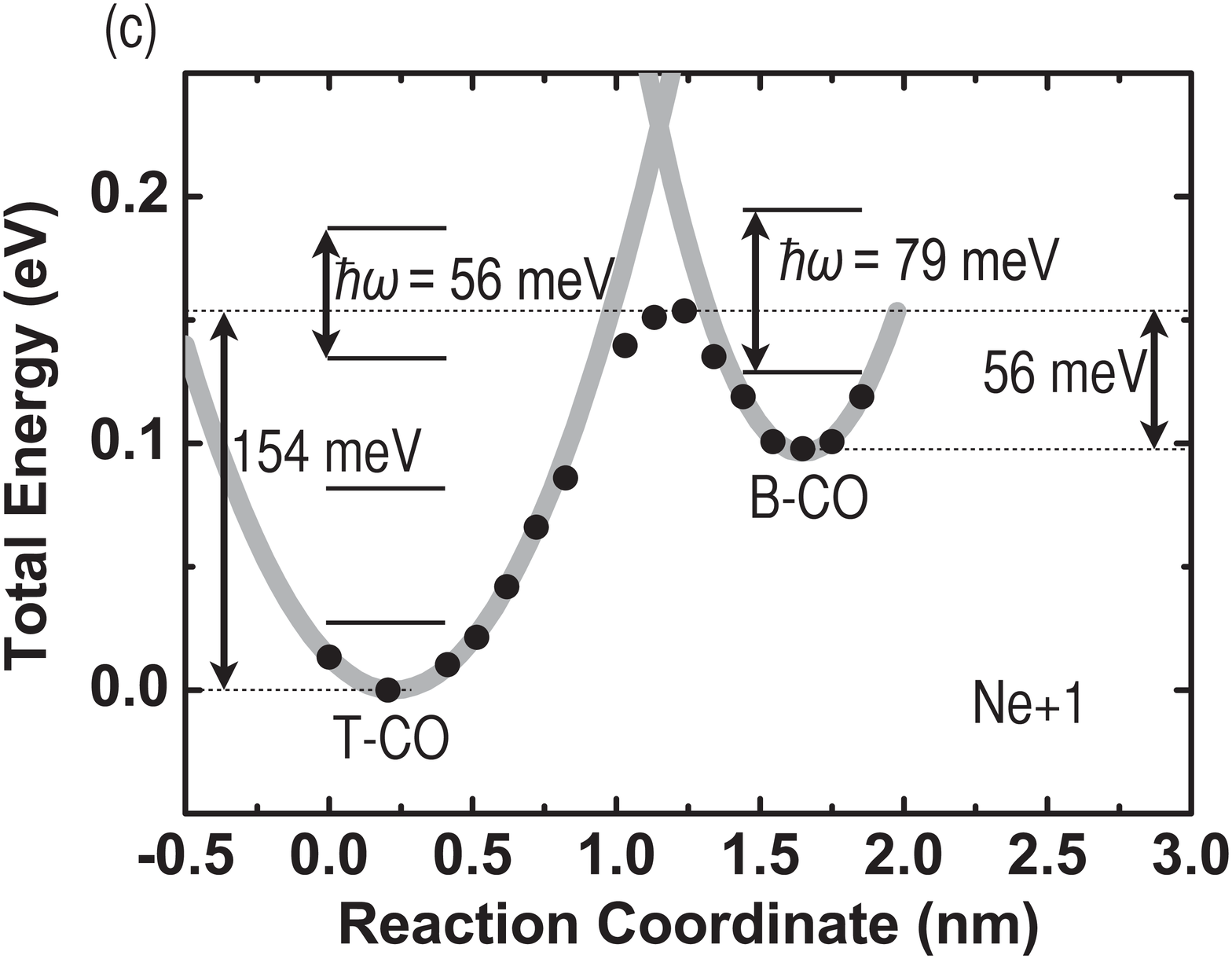}  
	 \caption{(a) Schematic model showing the transition from T-CO to B-CO on
Si(001).  The calculated potential curves on (b) the neutral slab $N_e$ and 
(c) the charged slab $N_e+1$ have different barriers. The calculation was done for a (2$\times$3)
unit cell with two fixed B-TO and one mobile. The reaction coordinate on the horizontal axis 
corresponds to the position of CO. 
The calculation involves an error of a few meV because the bulk atoms are not allowed to relax.
}
\end{figure}

\begin{figure}[htbp] 
   	 \centering
	 \includegraphics[width=12cm,clip]{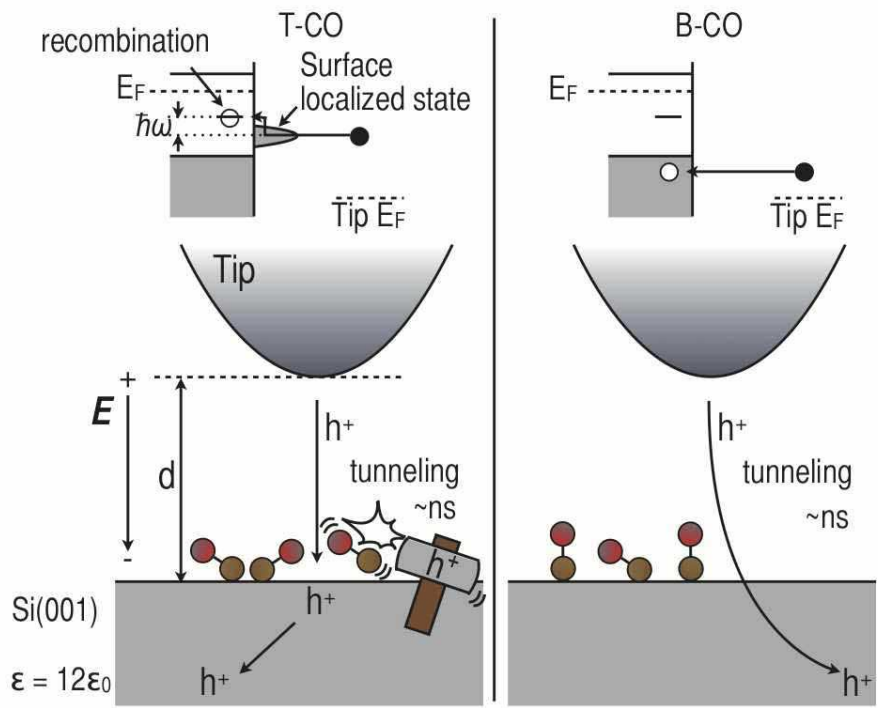} 
	 \caption{Top: schematic band diagrams for T-CO (left) and B-CO (right) in the case of negative sample bias voltage. 
The dashed lines among bands indicate the sample Fermi level ($E_F$).
Bottom: models of the impact of electron transfer as hole injection.  The hole tunnels to the mid-gap surface state near T-CO then goes into the bulk region, while at the B-CO site the hole tunnels to the substrate without being trapped near the surface region.}
	    \end{figure}

\end{document}